\documentclass[conference]{IEEEtran} 
\usepackage{amsmath}
\usepackage{cite}
\usepackage{amsfonts}
\usepackage{amssymb}
\usepackage{graphicx}
\usepackage{subfigure}
\usepackage{algorithm}
\usepackage{algpseudocode}
\usepackage{mathtools}

\usepackage{textcomp}
\usepackage[export]{adjustbox}
\usepackage{float}
\usepackage{booktabs}
\usepackage{multicol}
\usepackage{xparse}
\usepackage{color,soul}
\usepackage[bottom]{footmisc}
\usepackage[binary-units=true]{siunitx}
\DeclareSIUnit{\belmilliwatt}{Bm}
\DeclareSIUnit{\dBm}{\deci\belmilliwatt}
\DeclareSIUnit[per-mode=symbol,per-symbol=p]{\Bps}{\byte\per\second}
\sethlcolor{yellow}
\usepackage{algpseudocode}
\makeatletter
\def\BState{\State\hskip-\ALG@thistlm}
\makeatother

\begin{document}

\title{A Deep Reinforcement Learning Approach\\ for Improving Age of Information in Mission-Critical IoT
}

\author{\IEEEauthorblockN{Hossam Farag}
\IEEEauthorblockA{\textit{Department of Electronic Systems} \\
\textit{Aalborg University}\\
Copenhagen, Denmark \\
hmf@es.aau.dk}
\and
\IEEEauthorblockN{Mikael Gidlund}
\IEEEauthorblockA{\textit{Department of Information Systems and Technology} \\
\textit{Mid Sweden University}\\
Sundsvall, Sweden \\
mikael.gidlund@miun.se}
\and
\IEEEauthorblockN{\v{C}edomir Stefanovi\'{c}}
\IEEEauthorblockA{\textit{Department of Electronic Systems} \\
\textit{Aalborg University}\\
Copenhagen, Denmark  \\
cs@es.aau.dk}
}

	\maketitle
	\begin{abstract}
The emerging mission-critical Internet of Things (IoT) play a vital role in remote healthcare, haptic interaction, and industrial automation, where timely delivery of status updates is crucial. The Age of Information (AoI) is an effective metric to capture and evaluate information freshness at the destination. A system design based solely on the optimization of the average AoI might not be adequate to capture the requirements of mission-critical applications, since averaging eliminates the effects of extreme events. In this paper, we introduce a Deep Reinforcement Learning (DRL)-based algorithm to improve AoI in mission-critical IoT applications. The objective is to minimize an AoI-based metric consisting of the weighted sum of the average AoI and the probability of exceeding an AoI threshold. We utilize the actor-critic method to train the algorithm to achieve optimized scheduling policy to solve the formulated problem. The performance of our proposed method is evaluated in a simulated setup and the results show a significant improvement in terms of the average AoI and the AoI violation probability compared to the related-work.
	\end{abstract}
%%%%%%%%%%%%%%%%%%%%%%%%%%%%%%%%%%%%%%%%%%%%
%%%%%%%%%%%%%%%%%%%%%%%%%%%%%%%%%%%%%%%%%%%%	
\section{Introduction}
%%%%%%%%%%%%%%%%%%%%%%%%%%%%%%%%%%%%%%%%%%%%
Future Internet of Things (IoT) enables ubiquitous sensing and connected systems that will provide enhanced situational awareness, data driven decision analytics, and automated response without human intervention \cite{IoT}. Mission-critical IoT applications, such as first-responders monitoring, process automation and control, and intelligent transportation, are characterized by stringent communication requirements in terms of real-time  and reliable information delivery~\cite{urllc-iiot}. For such scenarios, packet delay cannot be considered a primary performance metric, but rather the freshness of the sensor data received at the destination side. Packets received with stale sensor data are useless to the destination, and may result in wrong actuation/control decisions as well. To this end, the Age of Information (AoI)~\cite{AoI}, is a relevant metric in quantifying the information freshness, which accounts for the time elapsed since the generation instant of the latest received update at the destination. The AoI depends both on the traffic generation pattern (frequent messages offering fresher information), as well as on the time spent by an update in the service. As a result, the metric is intrinsically different from classical performance criteria like throughput and latency, which focuses on a single packet and only captures its service time~\cite{AoI-minimum-urllc}.

Most previous works \cite{p1, p2, p3} focused on the analysis of the average AoI and peak AoI.
However, these two metrics of AoI can not fully characterize the precise performance of real-time status updates in mission-critical applications, as they do cannot account for extreme AoI events that occur with very low probabilities~\cite{AoI-vio}. Instead, the AoI distribution needs to be considered especially when dealing with time-critical IoT applications. Specifically, minimizing the violation probability that the AoI of each sensor node exceeds a predetermined age constraint is of great significance for guaranteeing information freshness in mission-critical networks. Therefore, the minimization problem should accounts for a combined AoI-based metric that considers the trade-off between minimizing the average AoI of each node and maintaining the violation probability at its minimum. This is a stochastic optimization problem with non-convex constraints that is known to be NP-hard even in deterministic settings. Deep Reinforcement Learning (DRL) is introduced as a viable way to solve NP-hard and non-convex problems~\cite{RL}, where an agent can be trained offline to choose the action that maximizes the system reward. 

In this paper, we introduce a DRL-based algorithm to minimize the average AoI and its threshold violation in mission-critical IoT networks.
We consider a remote monitoring and control scenario where a set of sensor nodes are deployed to transmit fresh updates to a central controller. 
We develop the problem as a NP-hard and non-convex problem aiming to minimize an objective function that constitutes the weighted sum of the average AoI and the probability of AoI threshold violation.
Then, we utilize DRL to solve the problem and produce the scheduling policy, i.e., which node is to be scheduled for transmitting fresh data at a given time instant that minimizes the formulated AoI-based metric.
The DRL algorithm is trained using the actor-critic algorithm with a non-zero probability of packet failure. Performance evaluations are carried out via simulations and the obtained results show that our proposed algorithm outperforms existing work in terms of the the average AoI and the AoI violation probability.

The remainder of the text is organized as follows. Section~II presents related work. Section~III introduces the proposed DRL method. Performance evaluations are given in Section~IV, followed by the conclusions in Section~V. 

%%%%%%%%%%%%%%%%%%%%%%5%%%%
\section{Related Work}
%%%%%%%%%%%%%%%%%%%%%%%%%%%
The problem of AoI minimization has been addressed in the literature considering various setups. The authors in~\cite{Aoi-m1} investigates the AoI minimization problem with minimum weighted sum of average peak ages for one-hop battery-free wireless sensor networks. In~\cite{Aoi-m2}, the authors consider the problem of scheduling real-time traffic to minimize the weighted average AoI under timely throughput constraint. An age-independent stationary randomized policy is proposed in~\cite{Aoi-m3} to minimize the long-term weighted AoI subject to a long-term average power constraint in fading multiple access channels. The authors of~\cite{Aoi-m4} also address the average AoI minimization problem under deterministic and random packet deadlines. In~\cite{Aoi-m5} the authors aim to minimize the long-term average AoI of  industrial IoT networks under average transmission power constraints at the source. Despite their interesting results, all the aforementioned works overlook the AoI violation probability in the formulated optimization problem. The work in~\cite{AoI-vio} introduces a theoretical basis for characterizing the violation probability of AoI in IoT systems under First-Come-First-Served (FCFS) discipline. The authors in~\cite{Aoi-m6} consider the minimization of the AoI violation probability in vehicular communication networks through an online learning approach. However, the proposed approach is not applicable in the considered industrial scenario of remote monitoring and control. 
%%%%%%%%%%%%%%%%%%%%%%%%%%%
	\begin{figure}[t!]
		\centering
		\includegraphics[width=0.9\linewidth]{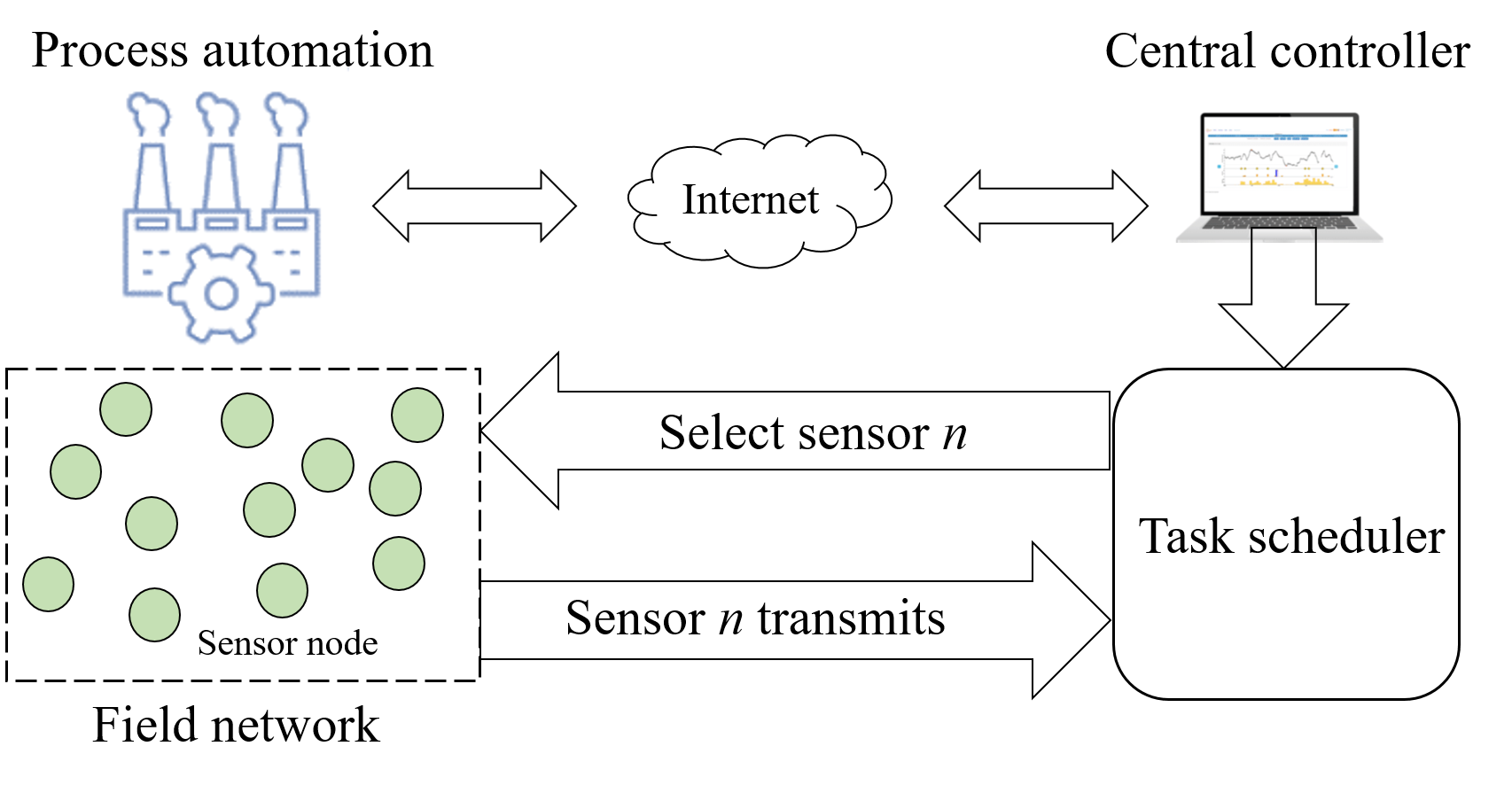}
		\caption{The remote monitoring system model. \label{sys-model}}
	\end{figure}
%%%%%%%%%%%%%%%%%%%%%%%%%%%%%%%%

%%%%%%%%%%%%%%%%%%%%%%%%%%%%%%%%
\section{The Proposed Method}
%%%%%%%%%%%%%%%%%%%%%%%%%%%%%%%%
In this work, we consider the remote monitoring system shown in Fig.\ref{sys-model}, in which $N$ sensor nodes are distributed at the field network side. The controller selects one node at a time to generate a fresh update and transmit it to a central controller.
The monitoring process is modelled as a set of tasks, where $i\in\{0, 1, ..., T-1\}$ denotes the task ID, and the scheduling policy implies that the controller selects one node $n\in \{0, 1, ..., N-1\}$ at a time to transmit its update at task ID $i$ at a rate $\lambda(i)$, i.e., $\lambda(i)$ represents the throughput achieved at the $i-th$ task. The variable $I_n(i)$ represents the output of the scheduler where $I_n(i)=1$ if sensor $n$ is selected to sample and transmit, and $I_n(i)=0$ otherwise. The sensors transmit updates with different sizes, where $L_n$ is the size of the transmitted packet by sensor $n$. at task number $i$. The packets are received successfully at the controller with probability $p\in(0,1]$, where $r_n(i)$ denotes the number of transmission attempts until sensor $n$ successfully transmits its packet at task $i$. Consequently, the age of the $n-th$ sensor's data at the controller can be expressed as
%%%%%%%%%%%%%%%%%%%%%%%%%%
\begin{equation}\label{AoI-n}
\begin{split}
A_n(i+1)&=I_n(i)\sum_{k=1}^{r_n(i)}\frac{L_n}{\lambda_k(i)}\\
&+(1-I_n(i))\left(A_n(i)+\sum_{m=1}^{N}I_m(i)\sum_{k=1}^{r_m(i)} \frac{L_m}{\lambda_k(i)}\right) 
\end{split}
\end{equation}
%%%%%%%%%%%%%%%%%%%%%%%%%%
where $\lambda_k(i)$ is the average rate at the $k-th$ transmission trial at the time of task $i$. The calculation of the age of sensor $n$ in \eqref{AoI-n} can be explained as follows. When sensor $n$ is selected by the controller at task $i$, i.e., $I_n(i)=1$, it is corresponding AoI at the controller side at task $i+1$ is equal to the time spent in transmitting its packet. When other sensor node $m\neq n$ is selected for transmission at task $i$, the AoI of sensor $n$ at task $i+1$ will be equal to the AoI at task $i$ plus the time spent in transmitting the packet from sensor $m$.

Our goal is to optimize the information freshness of the sensor nodes at the controller to improve performance of real-time status updates in mission-critical scenarios. Therefore, we formulate the optimization problem, in which the objective function considers both latency and reliability through a weighted sum of two AoI-based metrics as 
%%%%%%%%%%%%%%%%%%%%%%%%%%
\begin{equation} \label{opt-problem}
    \begin{split}
     \text{\textbf{minimize}}\,&\lim_{T\to\infty} \frac{1}{TN} \sum_{n=0}^{N-1}\sum_{i=0}^{T-1} A_n(i)\\
     &+\sum_{n=0}^{N-1}\delta_n\sum_{i=0}^{T-1}\Pr[A_n(i)>\beta_n]\\
     \text{\textbf{subject to}}\,\,&\eqref{AoI-n},\\
     &\sum_{n=0}^{N-1}I_n(i)=1,\,\forall i\in\{0, 1, ..., T-1\},\\ 
     & I_n(i)\in\{0,1\},\,\forall  n\in \{0, 1, ..., N-1\} ,
    \end{split}
\end{equation}
%%%%%%%%%%%%%%%%%%%%%%%%%%%
where $\beta_n$ is the AoI threshold of sensor $n$, where we assume that the sensors tolerate different AoI limits.
$\delta_n$ denotes the weight of sensor $n$ that quantifies the penalty of exceeding its specified AoI threshold.
The problem in \eqref{opt-problem} is a stochastic optimization problem with non-convex constraints that is known to be NP-hard. % and difficult to solve in polynomial time.
In the following, we introduce our proposed algorithm to solve the formulated optimization problem using DRL.

In our proposed method, each node (the agent) observes the state of the environment $s_i$, takes an action $a_i$ that leads to receiving a reward $R_i$ and the transition to a new state $s_{i+1}$.
The state $s_i$ includes the value of the AoI $A_n(i), \,\forall n,i$, the transmission time of the last packet denoted as $b(i-1)$ and the throughput that was achieved at the last $j$ tasks $\lambda (i-j+1), ..., \lambda (i+1)$.
The action $a(i)$ at the $i-th$ task represents the selection of the sensor $n$ that will transmit in the next task.
The set of actions is represented by a vector of probabilities\linebreak $\text{\textbf{P}}=\{p(0), p(2), ..., p(N-1)\}$ of length $N$, such that the action $a(i)$ is to select the sensor $n$ corresponds to the maximum element of $\text{\textbf{P}}$. The reward function $R_i$ is formulated as follows
%%%%%%%%%%%%%%%%%%%%%%%%%%%%%%%%%%
\begin{equation}\label{reward}
 R_i=\underbrace{-\sum_{n=0}^{N-1}A_n(i)}_{L_1}-\underbrace{\sum_{n=0}^{N-1}\delta_n \cdot \text{\textbf{1}}_{A_{n(i)>\beta_n}}}_{L_2}, 
\end{equation}
%%%%%%%%%%%%%%%%%%%%%%%%%%%%%%%%%%%
where $\text{\textbf{1}}_{(.)}$ is an indicator function. The first term $L_1$ in \eqref{reward} represents the sum of AoIs of all sensors at the end of the task $i$. The second term $L_2$ is the weighted sum of the penalties when exceeding the AoI thresholds. Therefore, the reward reflects the performance of each action with respect to the metric we need to optimize.
%%%%%%%%%%%%%%%%%%%%%%%%%%%
	\begin{figure}[t!]
		\centering
		\includegraphics[width=\linewidth]{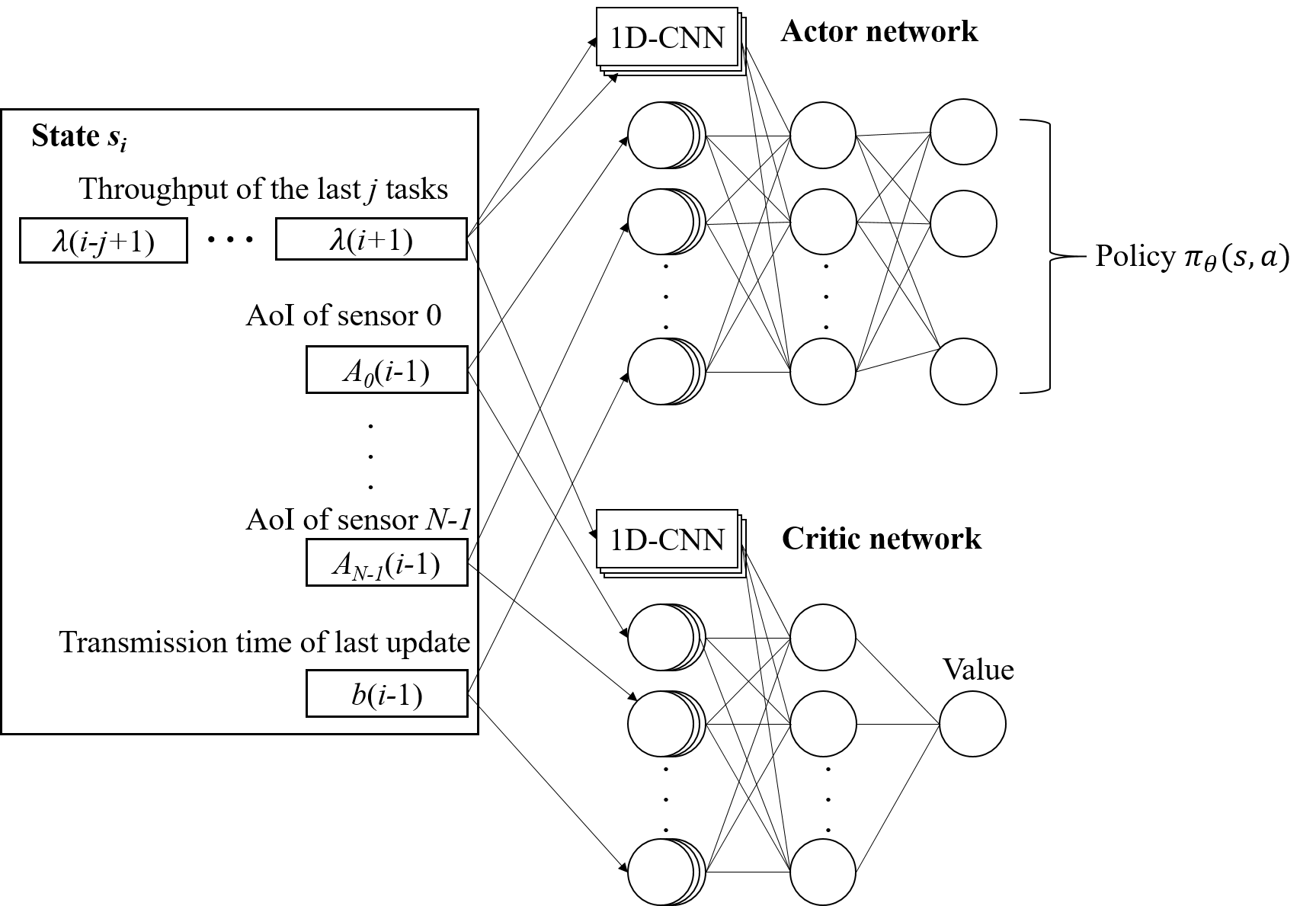}
		\caption{Structure of the A3C algorithm in our proposed method. \label{A3C-model}}
	\end{figure}
%%%%%%%%%%%%%%%%%%%%%%%%%%%%%%%%

In our proposed algorithm, first, the agent explores the environment through a training phase based on the Asynchronous Advantage Actor-Critic (A3C) algorithm~\cite{A3C}.
The structure of the A3C method is shown in Fig.~\ref{A3C-model} and is described in details as follows. The A3C is a model-free RL method that is used for directly updating a stochastic policy which runs multiple agents asynchronously with each agent having it’s own neural network. A3C algorithm is characterized by learning separate functions, one for the actor which is the parameterized policy $\pi_\theta(s_i,a_i)$ and the other one for the critic which is the value function $V^{\pi_\theta}(s_i,\theta)$, where $\theta$ is the policy parameter. The agent observes the environment through a set of metrics and feeds their values to the neural network as depicted in Fig.~\ref{A3C-model}. The critic takes the role of the value function and evaluates the performance of the actor, hereby helping with the estimation of the gradient to use for the actor’s updates. Hence, for each choice of the parameter $\theta$, we get the policy (probability over actions) $\pi_\theta(s_i,a_i)\leftarrow [0,1]$, where $\pi_\theta(s_i,a_i)$ is assumed to be differentiable over $\theta$, i.e., $\frac{\partial\pi_\theta(s_i,a_i)}{\partial\theta}$ exists. Gradient updates  are performed to find the parameters that lead to maximal rewards using the  policy gradient method \cite{grad}. The method  estimates the gradient of the cumulative discounted reward with respect to the policy parameter as 
%%%%%%%%%%%%%%%%%%%%%%%%%%%%%%%%%%
\begin{equation}\label{gradient}
\bigtriangledown E_{\pi_ \theta}\{\sum_{i=0}^{T-1}\sum_{j=0}^{T-1}\gamma^{i} R_{i+j}\}= E_{\pi_\theta}\{\bigtriangledown \text{log}\,\pi_\theta(s_i,a_i) D^{\pi_\theta}(s_i,a_i),   
\end{equation}
%%%%%%%%%%%%%%%%%%%%%%%%%%%%%%%%
where $\gamma\in [0,1]$ is the discount rate and $D^{\pi_\theta}(s_i,a_i)$ is the advantage function \cite{A3C} that captures how better an action is compared to the expected one selected according to the policy. The agent estimates $D^{\pi_\theta}(s_i,a_i)$ by sampling a trajectory of scheduling decisions and uses the empirically computed advantage as an unbiased estimate of $D^{\pi(\theta)}(s_i,a_i)$. Then, the learning parameter $\theta$ is updated as follows
%%%%%%%%%%%%%%%%%%%%%%%%%%%%%%%%
\begin{equation}
\theta  \leftarrow \theta +\alpha\sum_{i=0}^{T-1}\bigtriangledown\text{log}\,\pi_\theta(s_i,a_i) D^{\pi_\theta}(s_i,a_i),
\end{equation}
%%%%%%%%%%%%%%%%%%%%%%%%%%%%%%%%
where $\alpha$ is the learning rate. The advantage function $ D^{\pi_\theta}(s_i,a_i)$ is computed based on the estimation of the value function of the current state $v^{\pi_\theta}(s,\theta)$. The value of $v^{\pi_\theta}(s,\theta)$ is calculated by the critic network as the expected total reward starting at state $s$ and following the policy $\pi_\theta$. The critic network is trained based on the Temporal Difference (TD) method \cite{TD} by bootstrapping from the current estimate of the value function. Therefore, the update $\theta_v$ of the learning parameter is given as
%%%%%%%%%%%%%%%%%%%%%%%%%%%
\begin{equation}
    \theta_v \leftarrow \theta_v +\grave{\alpha}\sum_{i=0}^{T-1}\left(R_{i}+\gamma V^{\pi_\theta}(s_{i+1},\theta_v)-V^{\pi_\theta}(s_i,\theta_v)\right)^2,
\end{equation}
%%%%%%%%%%%%%%%%%%%%%%%%%%%
where $\grave{\alpha}$ is the learning rate of the critic network and $V^{\pi_\theta}$ is the estimate of $v^{\pi_\theta}$. Hence, for a given $(R_i,s_i,a_i,s_{i+1})$, the advantage function $D^{\pi_\theta}(s_i,a_i)$ is estimated as $R_{i}+\gamma V^{\pi_\theta}(s_{i+1},\theta_v)-V^{\pi_\theta}(s_i,\theta_v)$.

%%%%%%%%%%%%%%%%%%%%%%%%%%%%%%%%%%%%%%%%%%%%
\section{Performance Evaluation}
%%%%%%%%%%%%%%%%%%%%%%%%%%%%%%%%%%%%%%%%%%%
In this section, we first describe the simulation environment of our proposed method, then we introduce and discuss the simulation results. 

%%%%%%%%%%%%%%%%%%
\begin{table}
		\centering
		\caption{Evaluation parameters.}
		\label{t1}
		\begin{tabular}{ll}
			\toprule
			Parameter & Value \\
				\midrule
			Network size ($N$) & \num{10} nodes\\
			Packet length & [\SI{10}-\SI{100}{\byte}] \\
			AoI Threshold & [\SI{20}-\SI{200}{\milli\second}] \\
			Packet drop probability & \num{10}\%  \\
			$\alpha$ & \num{0.01}\\
			$\grave{\alpha}$ & \num{0.01}\\
			$\gamma$ & \num{100}\\
			$\rho$ & \num{5}\\
			\bottomrule
		\end{tabular}	
	\end{table}
%%%%%%%%%%%%%%%%%%%%%%%%
%%%%%%%%%%%%%%%%%%%%%%%%%
\begin{table}
    \centering
        \caption{Performance Comparison}
    \label{t-results}
\begin{tabular}{ |c|c|c|c| }
\hline
     & Proposed method & Benchmark \\ \hline
    Normalized objective function & 1.2 & 2.1 \\ \hline
    $P_{V0}$ & 0.07\% & 19.03\% \\ \hline
    $P_{V1}$ & 0.01\% & 8.55\% \\ \hline 
    $P_{V2}$ & 0.016\% & 10.12\% \\ \hline
    $P_{V3}$ & 0.01\% & 12.02\% \\ \hline 
    $P_{V4}$ & 0\% & 7.24\% \\ \hline
    $P_{V5}$ & 0\% & 4.24\% \\ \hline 
    $P_{V6}$ & 0\% & 3.57\% \\ \hline
    $P_{V7}$ & 0\% & 5.22\% \\ \hline 
    $P_{V8}$ & 0\% & 3.44\% \\ \hline
    $P_{V9}$ & 0\% & 4.30\% \\ \hline 
   Average AoI ($n=0$)& 11.02 & 26.34 \\ \hline
   Average AoI ($n=1$)& 18.25 & 33.66 \\ \hline
   Average AoI ($n=2$)& 20.11 & 40.73 \\ \hline
   Average AoI ($n=3$)& 28.39 & 44.18 \\ \hline
   Average AoI ($n=4$)& 30.64 & 51.22\\ \hline
   Average AoI ($n=5$)& 34.07 & 58.68\\ \hline
   Average AoI ($n=6$)& 30.89 & 61.48\\ \hline
   Average AoI ($n=7$)& 37.14 & 70.29\\ \hline
   Average AoI ($n=8$)& 40.01 & 81.98 \\ \hline
   Average AoI ($n=9$)& 39.71 & 89.53\\ \hline
    \end{tabular}
\end{table}
%%%%%%%%%%%%%%%%%%%%%%%%%%%
	\begin{figure*}[t!]
		\centering
		\includegraphics[width=\linewidth]{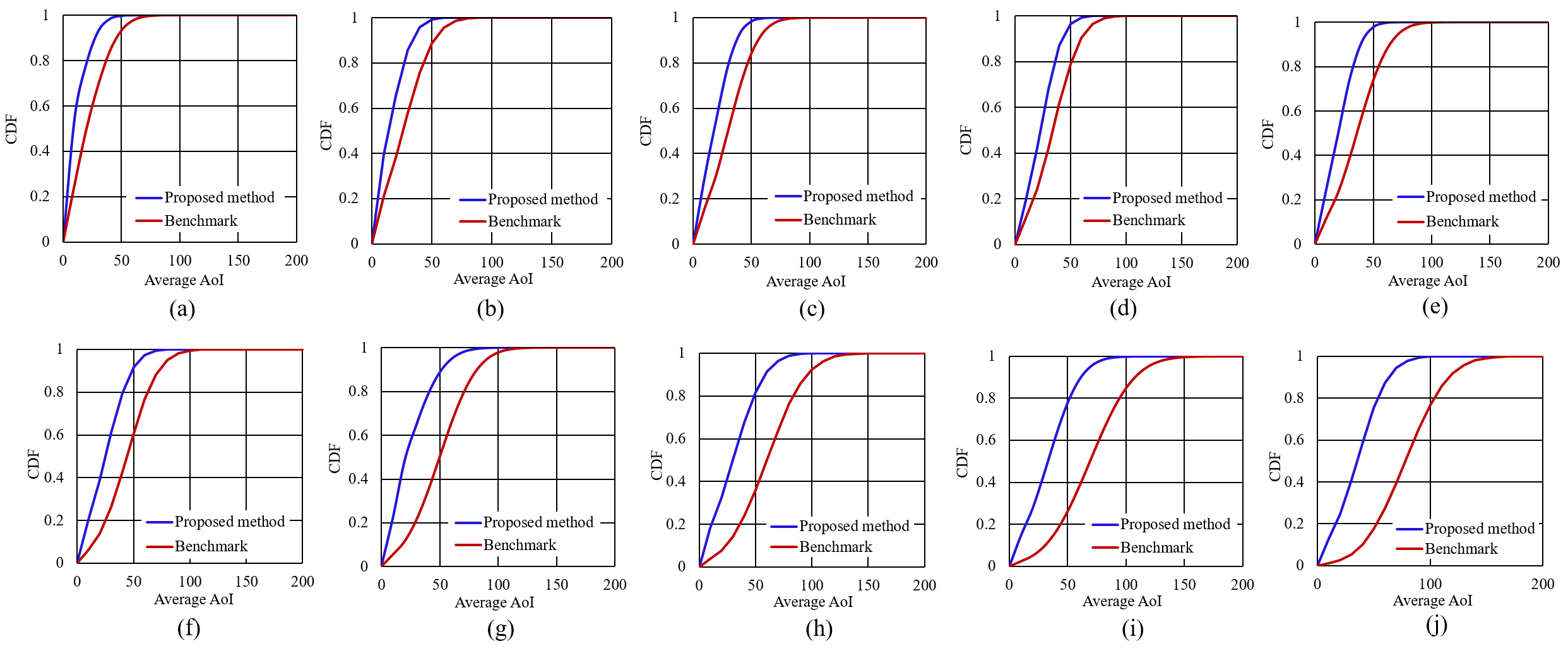}
		\caption{CDF of the AoI of each sensor; (a) $n=0$, (b) $n=1$, ..., (j) $n=9$ \label{CDF}}
	\end{figure*}
%%%%%%%%%%%%%%%%%%%%%%%%%%%%%%%%
%%%%%%%%%%%%%%%%%%%%%%%%%%%%
\subsection{Simulation Model}
%%%%%%%%%%%%%%%%%%%%%%%%%%%%%
 We evaluate the performance of our proposed method via MATLAB simulations using the parameters listed in Table~\ref{t1}.
 We consider a field network consists of 10 sensors (i.e., $N=10$) where each node generates packets of sizes 10 to 100~bytes (with a step size of 10). Each sensor node is attached with a different AoI threshold within the  range [\SI{20}-\SI{200}{\milli\second}] with a step size of \SI{20}{\milli\second}, where node $n=0$ has the strictest threshold and node $n=9$ has the loosest one. We set $\delta_n=1000\frac{N-n}{N}$, so that nodes with tighter AoI thresholds get more penalty. The actor-critic pair is implemented in each node (agent) with the same neural network structure. We use a 1D Convolutional Neural Network (1D-CNN) layer with 128 filters each of size 8 and stride 1, followed by a hidden layer consists of 256 neurons with a Rectified Linear Unit (ReLU) activation function \cite{relu}. The throughput of the last 10 tasks ($j=10$) is passed to the 1D-CNN, then its output along with the other inputs is aggregated in the hidden layer. Finally, the model has two sets of outputs, a single linear output for the value function and 10 terminals representing the output of the policy function based on the softmax approach. To ensure that each agent explores the action space adequately during training, we add an entropy regularization term \cite{A3C} with a weight $\rho$ to the actor’s update rule. This term encourages exploration by pushing $\theta$ in the direction of higher entropy. $\rho$ is set to 5 at the start of training and  decays until it reaches zero. The work proposed in \cite{Aoi-m2} is used as a benchmark to evaluate the effectiveness of our proposed method. In the referred work, a sensor $n$ is selected at task $i$ with a probability that is inversely proportional to its AoI threshold. 
%%%%%%%%%%%%%%%%%%%%%%%%%%%%%%%%
\subsection{Results and Discussion}
%%%%%%%%%%%%%%%%%%%%%%%%%%%%%%%%
Here we present the simulation results and the performance comparison with regard to the work in~\cite{Aoi-m2} (referred to as ``Benchmark" in the results). 
The obtained results are summarized in Table~\ref{t-results} and Fig.~\ref{CDF}. 

In Table~\ref{t-results}, we include three metrics for comparison, the normalized objective function in~\eqref{opt-problem}, the violation probability of each sensor node ($P_{Vn}=\Pr [A_n (i)>\beta_n]$, and the average AoI of each sensor node. From Table~\ref{t-results}, it can be observed that the proposed method achieves a 57\% lower value for the objective function compared to the Benchmark. This is mainly due to the proposed A3C learning algorithm which guides the scheduling policy to the minimum sum of both the average AoI and the violation probability. Moreover, Table~\ref{t-results} shows that our proposed method outperform the Benchmark in terms of the average AoI and the AoI violation probability for all the sensors, even for the ones that have strict deadlines. For instance, we can observe that  node 0 (i.e., $n=0$) exceeds its AoI threshold with a probability of 0.07\% which is~99\% lower than that of the same sensor using the method in the Benchmark. Furthermore, the proposed algorithm eliminates the AoI violations for nodes $n=4$ up to $n=9$, while the Benchmark continues to experience high threshold violation for all sensors, e.g., at least the nodes exceed their AoI threshold $4\%$ of the time. In the proposed method, the nodes learn how to respect the AoI violation threshold by giving weights that are directly proportional to the specified AoI limit, of which function is not defined by the Benchmark. 

In order to take a deeper look into the performance of the AoI, the plots in Fig.~\ref{CDF} show the Cumulative Distribution Function (CDF) of the AoI of each sensor for the two methods. From these plots, it can be clearly noticed that the proposed method achieves a reduced AoI all the time for all sensors compared to the Benchmark. The CDF performance of the proposed method also confirms the improvements achieved in the AoI violation probability given in Table~\ref{t-results}. As a summary, our results show that our proposed learning method guides the nodes to minimize the average AoI while maintaining low AoI violation probability as much as possible, which essential in mission-critical application within the IoT scenarios.

%%%%%%%%%%%%%%%%%%%%%%%%%%%%%%%%
\section{Conclusion}
%%%%%%%%%%%%%%%%%%%%
In this paper, we developed a DRL-based algorithm to improve the AoI performance of mission-critical IoT. We formulate an optimization problem to minimize the weighted sum of the average AoI and the probability of exceeding AoI threshold. We employed the A3C approach to train our algorithm and solve the formulated problem. Compared with existing work, simulation results proved the effectiveness of our proposed method with respect to the formulated AoI-based metric. An interesting direction for future work could be investigating the performance of the proposed method in large-scale networks, and also integrating a sampling process to the optimization problem and scheduling decision process.   
%%%%%%%%%%%%%%%%%%%%

\section*{Acknowledgement}

This paper has received funding from the European Union’s Horizon 2020 research and innovation programme under grant agreement No. 883315.

\end{document}